\documentclass{article}

\usepackage{times}
\usepackage{graphicx} % more modern
\usepackage{subfigure}

\usepackage{natbib}
\usepackage{amsmath}
\usepackage{amssymb}
\usepackage{algorithm}
\usepackage{algorithmic}
\usepackage{multirow}
\usepackage{booktabs}

\usepackage{hyperref}
\usepackage{bm}
\usepackage{url}
\usepackage{xspace}
\newcommand{\cfnade}{CF-NADE\xspace}

\newcommand{\reals}{\mathbb{R}}

\newcommand{\cost}{\mathcal{C}}
\newboolean{for_submission}
\setboolean{for_submission}{false}
\newcommand{\says}[3]{\ifthenelse{\boolean{for_submission}}{}{{\color{#3}#1 says: \emph{#2}\color{black}}\xspace}}

\usepackage[accepted]{icml2016}

\icmltitlerunning{A Neural Autoregressive Approach to Collaborative Filtering}

\begin{document} 
\twocolumn[
\icmltitle{A Neural Autoregressive Approach to Collaborative Filtering}

% It is OKAY to include author information, even for blind
% submissions: the style file will automatically remove it for you
% unless you've provided the [accepted] option to the icml2016
% package.
\icmlauthor{Yin Zheng}{yin.zheng@hulu.com}
%\icmladdress{HULU LLC, Beijing, 100084}
\icmlauthor{Bangsheng Tang}{bangsheng@hulu.com}
%\icmladdress{HULU LLC, Beijing, 100084}
%\icmlauthor{Cailiang Liu}{cailiang@hulu.com}
%\icmladdress{HULU LLC, Beijing, 100084}
\icmlauthor{Wenkui Ding}{wenkui.ding@hulu.com}
%\icmladdress{HULU LLC, Beijing, 100084}
\icmlauthor{Hanning Zhou}{eric.zhou@hulu.com}
\icmladdress{Hulu LLC., Beijing, 100084}
% You may provide any keywords that you 
% find helpful for describing your paper; these are used to populate 
% the "keywords" metadata in the PDF but will not be shown in the document
\icmlkeywords{boring formatting information, machine learning, ICML}

\vskip 0.3in
]
%\ericsays{blah blah blah}
%
%\bangshengsays{these annotations can be conveniently switched off}
%
%\wenkuisays{as we did last year}
%
%\yinsays{by turning off one boolean variable}

\begin{abstract} 

%This paper proposes \cfnade, a neural autoregressive architecture for collaborative filtering (CF) tasks. \cfnade is inspired by Restricted Boltzmann Machine (RBM) based CF model and Neural Autoregressive Distribution Estimator (NADE) model. In this paper, we first generalize NADE to \cfnade to deal with CF tasks. Then we describe how to improve the performance of \cfnade by adding rating-invariant information into the model and sharing parameters between different ratings. A factored version of \cfnade is also proposed to deal with large-scale dataset efficiently. Moreover, we take the ordinal nature of preference into consideration and propose an ordinal cost to optimize \cfnade. Finally, we extend \cfnade to a deep model with reasonable increase of computation complexity. Experimental results show that \cfnade outperforms the state of the art methods on MovieLens 1M, MovieLens 10M and Netflix datasets.

This paper proposes \cfnade, a neural autoregressive
  architecture for collaborative filtering (CF) tasks, which is
  inspired by the Restricted Boltzmann Machine (RBM) based CF model and
  the Neural Autoregressive Distribution Estimator (NADE). We first describe the basic \cfnade model for CF tasks. 
  Then we propose to improve the model by 
  sharing parameters between different ratings. A factored version of
  \cfnade is also proposed for better scalability. Furthermore,
  we take the ordinal nature of the preferences into consideration and propose an ordinal cost to optimize \cfnade, which shows superior performance. 
  Finally, \cfnade can be extended to a deep model, with only
  moderately increased computational complexity. Experimental results
  show that \cfnade with a single hidden layer beats all previous
  state-of-the-art methods on MovieLens 1M, MovieLens 10M, and Netflix
  datasets, and adding more hidden layers can further improve the performance.

\end{abstract} 

\section{Introduction}
\label{sec:intro}  

%Collaborative filtering (CF) is the method of predicting the \textit{preference} or \textit{rating} that a user would give to an item according to a user's 
%past behaviors (ratings or preferences) and the decisions made by other similar users. CF is one of the key methods behind recommender systems and is increasingly attracting attention with the rapid development of e-commercial and social network recently. The assumption behind CF is that people who have similar preferences in the past will have similar preferences in the future. A good CF algorithm can help users to find the products or services that suit their personalized taste efficiently. 

Collaborative filtering (CF) is a class of methods for predicting a
user's \emph{preference} or \emph{rating} of an item, based on his/her
previous preferences or ratings and decisions made by \emph{similar}
users. CF lies at the core of most recommender systems and has
attracted increasing attention along with the recent boom of
e-commerce and social network systems. The premise of CF is that a
person's preference does not change much over time. A good CF
algorithm helps a user discover products or services that suit his/her
taste efficiently.

%Generally speaking, a common distinction made in CF literature is between Memory-based CF and Model-based CF. Memory-based CF usually computes the similarities between users or items directly from the rating data, which are then used for recommendation. The explainatbility of the recommended results as well as the easy-to-implement nature of memory-based CF make it be used widely in early recommender systems~\citep{resnick1994grouplens}. However, the limitations of memory-based CF, such as the performance decreases when data gets sparser, make it unreliable and hence is not popular nowadays. 

Generally speaking, there is a dichotomy of CF methods: Memory-based CF and Model-based CF. Memory-based CF usually computes the similarities between users or items directly from the rating data, which are then used for recommendation.  The explainatbility of the recommended results as well as the easy-to-implement nature of memory-based CF ensured its popularity in early recommender
  systems~\citep{resnick1994grouplens}. However, memory-based CF has
  faded out due to its poor performance on real-life large-scale and sparse data.

Distinct from memory-based CF, model-based CF learns
  a model from historical data and then uses the model to predict
  preferences of users. The models are usually developed with machine
  learning algorithms, such as Bayesian networks, clustering models
  and latent semantic models. Complex preference patterns can be
  recognized by these models, allowing model-based CF to perform better for
   preference prediction tasks.  Among all these models,
  matrix factorization is most popular and successful, c.f.
  \citep{koren2009matrix,salakhutdinov2008bayesian,mackey2011divide,gopalan2013scalable}.

  With the recent development of deep
  learning~\citep{krizhevsky2012imagenet,szegedy2014going,he2015deep},
  neural network based CF, a subclass of model-based CF, has gained
  enormous attention. A prominent example is RBM-based CF
  (RBM-CF)~\citep{salakhutdinov2007restricted}. RBM-CF is a two-layer
  undirected generative graph model which generalizes Restricted
  Boltzmann Machine (RBM) to modeling the distribution of tabular
  data, such as user's ratings of movies. RBM-CF has shown its power
  in Netflix prize challenge. However, RBM-CF suffers from inaccuracy
  and impractically long training time, since training RBM-CF is
  intractable and one has to rely on variational approximation
  or MCMC sampling.

%Recently,  a good alternative to RBM  has been proposed by \citet{larochelle2011neural}. The so called  Neural Autoregressive Distribution Estimator (NADE) is a tractable distribution estimator for high dimensional binary vectors. NADE computes the conditional probabilities of each element given the other elements to its left in the binary vector, where all conditionals share the same parameters. The probability of the binary vector can then be obtained by taking the product of these conditionals. Unlike RBM, NADE does not incorporate any latent variable where expensive inference must be performed, hence it can be optimized efficiently by backpropagation. NADE together with its variants achieved competitive results on many machine learning tasks~\citep{larochelle2012neural,uria2013rnade,zheng14sup,Uria2013b,zheng2014neural,zheng15deep}.

  Recently, a good alternative to RBM has been proposed by
  \citet{larochelle2011neural}. The so-called Neural Autoregressive
  Distribution Estimator (NADE) is a tractable distribution estimator
  for high dimensional binary vectors. NADE computes the conditional
  probabilities of each element given the other elements to its left
  in the binary vector, where all conditionals share the same
  parameters. The probability of the binary vector can then be
  obtained by taking the product of these conditionals. Unlike RBM,
  NADE does not incorporate any latent variable where expensive
  inference is required, in constrast it can be optimized efficiently
  by backpropagation. NADE together with its variants achieved
  competitive results on many machine learning
  tasks~\citep{larochelle2012neural,uria2013rnade,zheng14sup,Uria2013b,zheng2014neural,zheng15deep}.

  In this paper, we propose a novel model-based CF approach named
  \cfnade, inspired by RBM-CF and NADE models. Specifically, we will
  show how to adapt NADE to CF tasks and describe how to improve the
  performance of \cfnade by encouraging the model to share parameters
  between different ratings. We also propose a factored version of
  \cfnade to deal with large-scale dataset efficiently. As
  \citet{phung2009ordinal} observed, preference usually has the
  \emph{ordinal nature}: if the true rating of an item by a user is 3
  stars in a 5-star scale, then predicting 4 stars is preferred to
  predicting 5 stars. We take this ordinal nature of preferences into
  consideration and propose an ordinal cost to optimize
  \cfnade. Moreover, we propose a deep version of \cfnade, which can
  be optimized efficiently. The performance of \cfnade is tested on
  $3$ real world benchmarks: MovieLens 1M, MovieLens 10M and Netflix
  dataset. Experimental results show that \cfnade outperforms all previous
  state-of-the-art methods.

\section{Related Work}
\label{sec:related_work}

As mentioned previously, some of the most successful model-based CF
methods are based on matrix factorization (MF) techniques, where a
prevalent assumption is that the partially observed matrix is of low
rank. In general, MF characterizes both users and items by
vectors of latent factors, where the number of factors is much smaller
than the number of users or items, and the correlation between user
and item factor vectors are used for recommendation
tasks. Specifically, \citet{billsus1998learning} proposed to apply
Singular Value Decomposition (SVD) to CF tasks, which is an early work
on MF-based CF. Bias MF \cite{koren2009matrix} is proposed to improve
the performance of SVD by introducing systematic biases associated
with users and items. \citet{mnih2007probabilistic} extended MF to a
probabilistic linear model with Gaussian noise referred to as
Probabilistic Matrix Factorization (PMF), and showed that PMF
performed better than SVD. \citet{salakhutdinov2008bayesian} proposed
a Bayesian treatment of PMF, which can be trained efficiently by MCMC
methods. Along this line, \citet{lawrence2009non} proposed a non-linear
PMF using Gaussian process latent variable models. There are other
MF-based CF methods such as
\citep{rennie2005fast,mackey2011divide}. Recently, Poisson Matrix
Factorization~\citep{gopalan2014content,gopalan2014bayesian,gopalan2013scalable}
was proposed, replacing Gaussian assumption of PMF by Poisson
distribution. \citet{lee2013local} extended the low-rank assumption by
embedding locality into MF models and proposed Local Low-Rank Matrix
Approximation (LLORMA) method, which achieved impressive performance on several public benchmarks.

%The accuracy and good scalability lead MF to be popular \cite{takacs2008investigation}. 

Another line of model-based CF is based on neural networks. With the
tremendous success of deep
learning~\citep{krizhevsky2012imagenet,szegedy2014going,he2015deep},
neural networks have found profound applications in CF
tasks. \citet{salakhutdinov2007restricted} proposed a variant of
Restricted Boltzmann Machine (RBM) for CF tasks, which is successfully
applied in Netflix prize
challenge~\citep{bennett2007netflix}. Recently,
\citet{sedhain2015autorec} proposed AutoRec, an autoencoder-based CF
model, which achieved the state-of-the-art performance on some
benchmarks. RBM-CF~\citep{salakhutdinov2007restricted} and
AutoRec~\citep{sedhain2015autorec} are common in that both of them
build different models for different users, where all these models
share the parameters.  \citet{phung2009ordinal} proposed to apply
Boltzmann Machine (BM) on CF tasks, which extends RBM-CF by
integrating the correlation between users and between
items. \citet{phung2009ordinal} also extended the standard BM model so
as to exploit the ordinal nature of ratings. Recently,
\citet{dziugaite2015neural} proposed Neural Network Matrix
Factorization (NNMF), where the inner product between the vectors of
users and items in MF is replaced by a feed-forward neural
network. However, NNMF does not produce convincing results on
benchmarks.

Our proposed method \cfnade, can be generally categorized as a neural
network based CF method. \cfnade bears some similarities with
NADE~\citep{larochelle2011neural} in that both model vectors with
neural autoregressive architectures. The crucial difference between
\cfnade and NADE is that \cfnade is designed to deal with vectors of
variable length, while NADE can only deal with binary vectors of fixed
length. Though DocNADE~\citep{larochelle2012neural} does take inputs
with various lengths, it is designed to model unordered sets of words
where each element of the input corresponds to a word, while \cfnade
models user rating vectors, where each element corresponds to the
rating to a specific item.

%\ericsays{Our proposed method \cfnade, can be generally categorized as a neural network based CF method. \cfnade bears some similarities with NADE~\citep{larochelle2011neural} in that both model vectors with a neural autoregressive architecture. The crucial difference between \cfnade and NADE is that \cfnade is designed to deal with vectors of variable length, while NADE can only deal with binary vectors of fixed length. Another extension of NADE, DocNADE~\citep{larochelle2012neural}, also takes variable length input. However, DocNADE is designed to model documents represented by Bag of Words (BOW) features, while \cfnade focuses on modeling vectors. }

\section{NADE for Collaborative Filtering}
\label{sec:cfnade}
%This section will describe \cfnade, a NADE based model for CF tasks. Specifically, we first describe the basic model of \cfnade in Section~\ref{sec:model}. Then we introduce two ways to improve the performance of \cfnade in Section~\ref{sec:inputmask} and Section~\ref{sec:accu}, respectively. At last, a factored version of \cfnade  is described to deal with large-scale dataset in Section~\ref{sec:factored}.

This section devotes to \cfnade, a NADE-based model for CF
tasks. Specifically, we describe the basic model of \cfnade in
Section~\ref{sec:model}, and propose to improve \cfnade by
sharing parameters between different ratings in
Section~\ref{sec:accu}. At last, a factored version of \cfnade is
described in Section~\ref{sec:factored} to deal with large-scale
datasets.

%by decomposing large connection matrix by a product of two low-rank matrices.

\subsection{The Model} 
\label{sec:model}

Suppose that there are $M$ items, $N$ users, and the ratings are
integers from $1$ to $K$ ($K$-star scale). One practical and prevalent
assumption in CF literature is that a user usually rated $D$
items\footnote{$D$ might vary between different users}, where
$D \ll M$. To tackle sparsity, similar to
RBM-CF~\citep{salakhutdinov2007restricted}, we use a different \cfnade
model for each user and all these models share the same
parameters. Specifically, all models have the same number of hidden
units, but a user-specific model only has $D$ visible units if the user
only rated $D$ items. Thus, each \cfnade has only one single training
case, which is a vector of ratings that a user gave to his/her viewed
items, but all the weights and biases of these \cfnade 's are tied.

In this paper, we denote the training case for user $u$ as
$\mathbf{r}^{u} = (r^{u}_{m_{o_1}}, r^{u}_{m_{o_2}}, \ldots,
r^{u}_{m_{o_D}})$,
where $o$ is a $D$-tuple in the set of permutations of
$(1,2,\ldots,D)$ which serves as an ordering of the rated items
$m_i\in \{1,2,\ldots,M\}$, and $r^{u}_{m_{o_i}} \in \{1,2,\ldots,K\}$
denotes the rating that the user gave to item $m_{o_i}$. For
simplicity, we will omit the index $u$ of ${\bf r}^u$, and focus on a
single user-specific \cfnade in the rest of the paper.

%\bangshengsays{ Suppose that there are $M$ items, $N$ users, and the
%  ratings are integers ranging from $1$ to $K$ ($K$-star scale). A
%  practical assumption prevalent in CF literature is that the ratings
%  are sparse, in that a user $u$ only rates $D_u$ items, where
%  $D_u\ll M$. To tackle sparsity, similar to
%  RBM-CF~\citep{salakhutdinov2007restricted}, we use a different
%  \cfnade model for each user and all these models share the same
%  parameters. Specifically, these \cfnade models have the same number
%  of hidden units, but the \cfnade model for $u$ has only $D_u$
%  visible units. Thus, each \cfnade has one single training case,
%  which is a vector of ratings that a user $u$ gives to his/her viewed
%  items, but all the weights and biases of these \cfnade 's are tied.
%}

%\bangshengsays{ In this paper, we denote the training case for user
%  $u$ as
%  $\mathbf{r}^{u} = (r^{u}_{m_{o_1}}, r^{u}_{m_{o_2}}, \ldots,
%  r^{u}_{m_{o_{D_u}}})$,
%  where $o$ is a permutation of $\{1,2,\ldots,D_u\}$ which serves as
%  the ordering of the rated items $m_i\in \{1,2,\ldots,M\}$, and
%  $r^{u}_{m_{o_i}} \in \{1,2,\ldots,K\}$ denotes the rating that $u$
%  gives to item $m_{o_i}$. For simplicity, we will omit the index $u$
%  of ${\bf r}^u$ and $D_u$, and focus on a single user-specific
%  \cfnade in the rest of the paper. }

\cfnade models the probability of the rating vector $\bf r$ by the chain rule as:
\begin{equation}
    p\left({\bf r}\right) = \prod_{i=1}^{D} p\left(r_{m_{o_i}}|{\bf r}_{m_{o_{<i}}}\right)
    \label{eqn:chain_rule}
\end{equation}
where ${\bf r}_{m_{o_{<i}}} = (r_{m_{o_1}}, r_{m_{o_2}}, \ldots, r_{m_{o_{i-1}}})$ denotes the first $i-1$ elements of $\bf r$ indexed by $o$.

Similar to NADE~\citep{larochelle2011neural}, \cfnade models the
conditionals in Equation~\ref{eqn:chain_rule} with neural networks. To
compute the conditionals in Equation~\ref{eqn:chain_rule}, \cfnade
first computes the hidden representation of dimension $H$ given
${\bf r}_{m_{o_{<i}}}$ as follows:
\begin{equation}
\mathbf{h}\left ( {\bf r}_{m_{o_{<i}}} \right ) = {\bf g}\left( \mathbf{c}+\sum_{j<i}\mathbf{W}^{r_{m_{o_j}}}_{:,m_{o_j}} \right ) 
\label{eqn:cfnade_hidden}
\end{equation}
where $\bf g(\cdot)$ is the activation function, such as $\tanh(x) =
\frac{\exp(x)-\exp(-x)}{\exp(x)+\exp(-x)}$, ${\bf W}^k\in \reals^{H\times M }$ is the connection
matrix associated with rating $k$, ${\bf W}_{:,j}^k\in \reals^{H}$ is the $j$\textsuperscript{th} column of  ${\bf W}^k$ and $W^{k}_{i,j}$ is an interaction
parameter between the $i$\textsuperscript{th} hidden unit and item
$j$ with rating $k$,  ${\bf c}\in \reals^H$ is the bias term.

Then the conditionals in Equation~\ref{eqn:chain_rule} could be modeled as:
\begin{equation}
p\left ( r_{m_{o_i}}=k|{\bf r}_{m_{o_{<i}}} \right ) = \frac{\exp \left(s_{m_{o_i}}^{k}\left({\bf r}_{m_{o_{<i}}}\right)\right)}{\sum_{k'=1}^K\exp \left(s_{m_{o_i}}^{k'}\left({\bf r}_{m_{o_{<i}}}\right)\right)}
\label{eqn:cfnade_softmax} 
\end{equation}
where $s_{m_{o_i}}^{k}({\bf r}_{m_{o_{<i}}})$ is the score indicating the preference that the user gave rating $k$ for item $m_{o_i}$ given the previous ratings ${\bf r}_{m_{o_{<i}}}$, and $s_{m_{o_i}}^{k}({\bf r}_{m_{o_{<i}}})$ is computed as,
\begin{equation}
s_{m_{o_i}}^{k}\left({\bf r}_{m_{o_{<i}}}\right) =b^{k}_{m_{o_i}} +\mathbf{V}^{k}_{m_{o_i},:}\mathbf{h}\left ( {\bf r}_{m_{o_{<i}}} \right )
\label{eqn:score}
\end{equation}
where ${\bf V}^{k}\in \reals^{M \times H}$ and  ${\bf b}^{k} \in \reals^{M}$ are the connection matrix and the bias term associated with rating $k$, respectively. 

\cfnade is optimized for minimum negative log-likelihood of
$p\left({\bf r}\right)$ (Equation~\eqref{eqn:cfnade_softmax}),
\begin{equation}
    -\log p\left({\bf r}\right) = -\sum_{i=1}^{D}\log p\left(r_{m_{o_i}}|{\bf r}_{m_{o_{<i}}}\right) 
    \label{eqn:nll_cost}
\end{equation}
averaged over all the training cases. As in NADE, the ordering $o$ in
\cfnade must be predefined and fixed during training for each
user. Ideally, the ordering should follow the timestamps when the user
gave the ratings. In practice, we find that a ordering that is
randomly drawn from the set of permutations of $(1,2,\ldots,D^u)$ for
each user $u$ yields good results. As \citet{Uria2013b} observed, we can think of the models trained with different orderings as different instantiations of \cfnade for the same user.
Section~\ref{sec:deepcfnade} will discuss how to (virtually) train a factorial number of \cfnade 's with different orderings simultaneously, which is the key to extend \cfnade to a deep model efficiently.

%\bangshengsays{ averaged over all training cases. As in NADE, the
%  ordering $o$ in \cfnade should be predefined and fixed during
%  training for each user. Ideally, we could define the ordering as the
%  timestamps when the user gave the ratings. However, we find that a
%  random permutation for each user $u$ yields good results for
%  practical use. As \citet{Uria2013b} observed, we can think of the
%  models trained with different orderings as different instantiations
%  of \cfnade for the same user. Section~\ref{sec:deepcfnade} will
%  discuss how to train a factorial number of \cfnade 's with different
%  orderings simultaneously, which is the key to extend \cfnade to a
%  deep model.}

%Optimizing the training objective requires that 

Once the model is trained, given a user's past behavior $\mathbf{r} =
(r_{m_{o_1}}, r_{m_{o_2}}, \ldots, r_{m_{o_D}})$, the user's rating of
a new item $m^*$ can be predicted as 
\begin{equation}
{\hat r}_{m^*} = \mathbb{E}_{p\left ( r_{m^{*}}=k|{\bf r} \right )}\left[k\right]
\end{equation}
where conditional $p\left ( r_{m^{*}}=k|{\bf r} \right )$ are computed by Equation~\ref{eqn:cfnade_softmax} along with the hidden representation $\mathbf{h}\left ( {\bf r} \right )$ and score $s_{m^*}^{k}\left({\bf r}\right)$ computed as 
\begin{eqnarray}
s_{m^*}^{k}\left({\bf r}\right) &=&b^{k}_{m^*} +\mathbf{V}^{k}_{m^*,:}\mathbf{h}\left ( {\bf r} \right )\\
\mathbf{h}\left ( {\bf r} \right ) &=& {\bf g}\left( \mathbf{c}+\sum_{j=1}^D\mathbf{W}^{r_{m_{o_j}}}_{:,m_{o_j}} \right ) .
\end{eqnarray}

\subsection{Sharing Parameters Between Different Ratings}
\label{sec:accu}

In Equations~\ref{eqn:cfnade_hidden} and \ref{eqn:score}, the
connection matrices ${\bf W}^k$, ${\bf V}^k$ and the bias ${\bf b}^k$
are different for different ratings $k$'s. In other words, \cfnade
uses different parameters for different ratings. In practice, for a
specific item, some ratings can be much more often observed than
others.  As a result, parameters associated with a rare rating might
not be sufficiently optimized. To alleviate this problem, we propose
to share parameters between different ratings of the same item.

Particularly, we propose to compute the hidden representation $\mathbf{h} ( {\bf r}_{m_{o_{<i}}} )$ as follows:
\begin{equation}
  \mathbf{h}\left ( {\bf r}_{m_{o_{<i}}} \right ) = {\bf g}\left( \mathbf{c}+\sum_{j<i}\sum_{k=1}^{r_{m_{o_j}}}\mathbf{W}_{:,m_{o_j}}^{k} \right ) \label{eqn:cfnade_hidden_shared}  
\end{equation}
Note that, given an item $m_{o_j}$ rated $ r_{m_{o_j}}$ by the user,
$\mathbf{h}( {\bf r}_{m_{o_{<i}}} )$ depends on all the weights
${\bf W}^{k}$, $\forall k \le r_{m_{o_j}}$. Thus,
Equation~\ref{eqn:cfnade_hidden_shared} encourages a solution that
${\bf W}^{t}$ is utilized by all the ratings $k$, $\forall k \ge t$.

%not only the weights $\mathbf{W}_{:,m_{o_j},r_{m_{o_j}}}$ where $j<i$ as Equation~\ref{eqn:cfnade_hidden} does, but all the weights  $\mathbf{W}_{:,m_{o_j},k}$ where $j<i$ and $k \le r_{m_{o_j}}$. Hence, given a item $m_i$, $\mathbf{W}_{:,m_i,k}$ is shared by all ratings $j$ where $k\le j$, not only rating $k$.  

%We denote the \cfnade model with this kind of hidden representation as \cfnadein.

Similarly, the score $s_{m_{o_i}}^{k}({\bf r}_{m_{o_{<i}}})$ in
Equation~\ref{eqn:cfnade_softmax} is adjusted as
\begin{equation}
s_{m_{o_i}}^{k}\left({\bf r}_{m_{o_{<i}}}\right) =\sum_{j\le k} \left(b^{j}_{m_{o_i}} +\mathbf{V}^{j}_{m_{o_i},:}\mathbf{h}\left ( {\bf r}_{m_{o_{<i}}} \right )\right)
    \label{eqn:score_shared}
\end{equation}
where $\mathbf{V}^j$ and ${\bf b}^{j}$ are shared by the rating $k$, where $k \ge j$.

Sharing parameters between different ratings can again be understood
as a kind of regularization, which encourages the model to use as many
parameters as possible to explain the data. Experimental results in
Section~\ref{sec:variants} confirm the advantage of this
regularization.

%Equation~\ref{eqn:cfnade_hidden_shared} and
%\ref{eqn:score_shared} can be interpreted as a kind of regularization which encourages the parameters of the 
%model to be shared. We will compare the performance of 

\subsection{Dealing with Large-Scale Datasets}
\label{sec:factored}
One disadvantage of \cfnade we have described so far is that the
parameterization of ${\bf W}^k \in \reals^{H\times M}$ and
${\bf V}^k \in \reals^{M\times H}$, where $k$ ranges from $1$ to $K$,
will result in too many free parameters, especially when dealing with
massive datasets. For example, for the Netflix
dataset~\citep{bennett2007netflix}, when $H=500$, the number of free
parameters by ${\bf W}^k$ and ${\bf V}^k$ would be around $89$
million\footnote{Netflix dataset contains $17770$ movies ($M=17770$)
  and the ratings are $5$-star scale ($K=5$). Thus, the number of free
  parameters from ${\bf W}^k$ and ${\bf V}^k$ is
  $88850000=2\times 17770\times 5\times 500$.}. Although severe
overfitting can be avoided by proper weight-decay or
dropout~\citep{srivastava2014dropout}, learning such a huge network
would still be problematic.
% , especially when scaling up the number of items or hidden units.

Inspired by RBM-CF~\citep{salakhutdinov2007restricted} and
FixationNADE~\citep{zheng2014neural}, we propose to address this
problem by factorizing ${\bf W}^k$ and ${\bf V}^k$ into products of
two lower-rank matrices. Particularly,
\begin{eqnarray}
W_{i,m}^k &=& \sum_{j=1}^J B_{i,j}A_{j,m}^k\label{eqn:fact_W}\\
V_{m,i}^k &=& \sum_{j=1}^J P_{m,j}^k Q_{j,i}\label{eqn:fact_V}
\end{eqnarray}
where ${\bf A}^k\in \reals^{J\times M}$,
${\bf P}^k\in \reals^{M\times J}$, ${\bf B}\in \reals^{H\times J}$ and
${\bf Q}\in \reals^{J\times H}$ are lower-rank matrices with $J\ll H$
and $J\ll M$. For example, by setting $J=50$, the number of free
parameters for $\bf W$ and $\bf V$ decreases from $89$ million to
about $9$ million. In our experiments, this factored version of \cfnade
will be applied on large-scale datasets.

\section{Traing \cfnade with Ordinal Cost}
\label{sec:cost}

%Maximum likelihood training of \cfnade could be performed by minimizing the negative log-likelihood cost:
%\begin{equation}
%    \cost_{nll} = -\sum_{i=1}^{D}\log p\left(r_{m_{o_i}}|{\bf r}_{m_{o_{<i}}}\right) 
%    \label{eqn:nll_cost}
%\end{equation}
%where $p\left(r_{m_{o_i}}|{\bf r}_{m_{o_{<i}}}\right)$ is defined by Equation~\ref{eqn:cfnade_softmax}.

%As mentioned in Section~\ref{sec:model}, 
\cfnade can be trained by minimizing the negative log-likelihood based
on conditionals defined by Equation~\ref{eqn:cfnade_softmax}. To go
one step further, following \citet{phung2009ordinal}, we take the
\emph{ordinal nature} of a user's preference into consideration. That
is, if a user rated an item $k$, the preference of the user
to the ratings from $1$ to $k$ should increase monotonically and the
preference to the ratings from $k$ to $K$ should decrease
monotonically. The basic \cfnade treats different ratings as separate
labels, leaving the ordinal information not captured. Here we describe
how to equip \cfnade with an ordinal cost.

Formally, suppose $r_{m_{o_i}}=k$, the ranking of preferences over all
the possible ratings under the ordinal assumption can be expressed as:
\begin{eqnarray}
k \succ k-1 \succ &\ldots& \succ 1\label{eqn:ordinal_down} \\
k \succ k+1 \succ &\ldots& \succ K
\label{eqn:ordinal_up}
\end{eqnarray}
where $k \succ k-1$ denotes the preference of rating $k$ over $k-1$,
$k\in \{1,2,\ldots,K\}$\footnote{Equation~\ref{eqn:ordinal_down} is
  omitted if the true rating is $1$; likewise,
  Equation~\ref{eqn:ordinal_up} is omitted if the true rating is $K$.}.
Two rankings of ratings, ${\bf y}^{\textup{down}}$ and
${\bf y}^{\textup{up}}$, can be induced by
Equation~\ref{eqn:ordinal_down} and \ref{eqn:ordinal_up}:
\begin{eqnarray}
{\bf y}^{\textup{down}} &=& \left(k, k-1,\ldots , 1\right)\label{eqn:y_down} \\
{\bf y}^{\textup{up}} &=& \left(k, k+1,\ldots , K\right)\label{eqn:y_up}
\end{eqnarray}

Note that maximizing the conditional $p( r_{m_{o_i}}=k|{\bf r}_{m_{o_{<i}}})$ in Equation~\ref{eqn:cfnade_softmax} only ensures that the probability of rating $k$ is the largest among all possible ratings. To capture the ordinal nature induced by Equations~\ref{eqn:ordinal_down} and \ref{eqn:ordinal_up}, we propose to compute the conditional $p( r_{m_{o_i}}=k|{\bf r}_{m_{o_{<i}}})$ as
\begin{multline}
p\left( r_{m_{o_i}}=k|{\bf r}_{m_{o_{<i}}}\right)=\\
\prod_{j=k}^1 \frac{\exp(s_{m_{o_i}}^{j})}{\sum_{t=1}^j \exp(s_{m_{o_i}}^{t})}\prod_{j=k}^{K} \frac{\exp(s_{m_{o_i}}^{j})}{\sum_{t=j}^K \exp(s_{m_{o_i}}^{t})}
\label{eqn:ordinal_conditional}
\end{multline}
where $s_{m_{o_i}}^{j}$ is a shorthand for the score $s_{m_{o_i}}^{j}({\bf r}_{m_{o_{<i}}})$ introduced in Section~\ref{sec:model}, which indicates the preference to rating $k$ of item $m_{o_i}$ given the previous context ${\bf r}_{m_{o_{<i}}}$. 

Both two products in Equation~\ref{eqn:ordinal_conditional} can be
interpreted as the \emph{likelihood loss} introduced
in~\citep{xia2008listwise} in the context of \emph{Listwise Learning
  To Rank} problem. Actually, from the perspective of
learning-to-rank, \cfnade acts as a ranking function which produces
rankings of ratings based on previous ratings, where
$s_{m_{o_i}}^{j}({\bf r}_{m_{o_{<i}}})$ corresponds to the score
function in~\citep{xia2008listwise} and the rankings,
${\bf y}^{\textup{down}}$ and ${\bf y}^{\textup{up}}$, corresponds to
true rankings that we would like \cfnade to fit. Thus, the conditional
computed by Equation~\ref{eqn:ordinal_conditional} is actually the
conditional distribution of the rankings ${\bf y}^{\textup{down}}$ and
${\bf y}^{\textup{up}}$ given previous $i-1$ ratings. Put differently, the ranking loss in Equation~\ref{eqn:ordinal_conditional}
is defined on the ratings, while other learning-to-rank based CF methods, such as \citep{shi2010list}, are on items, which is the crucial difference.

For the rest of the paper, we denote the negative log-likelihood based
on the conditionals computed by Equation~\ref{eqn:ordinal_conditional}
as \textbf{ordinal cost} $\cost_{\textup{ord}}$, and denote the
negative log-likelihood based on Equation~\ref{eqn:cfnade_softmax} as
\textbf{regular cost} $\cost_{\textup{reg}}$. The final cost to
optimize the model is then defined as
\begin{equation}
    \cost_{\textup{hybrid}} = (1-\lambda)\cost_{\textup{reg}} + \lambda\cost_{\textup{ord}}
    \label{eqn:hybrid_cost}
\end{equation}
where $\lambda$ is the hyperparameter to determine the weight of
$\cost_{\textup{ord}}$. The impact of the hyperparameter $\lambda$ on
the performance of \cfnade is discussed in Section~\ref{sec:variants}.

% TODO: difference with original NADE
% 1. D is different
% 2. The order of the ratings is shuffled
% 3. multinomial observations, not binary or real
% TODO: difference with DocNADE
% 1. Not BOW representation
% 2. 
% TODO: the introduce of RBM based CF, how do deal with missing ratings, etc.

\section{Extending \cfnade to a Deep Model}
\label{sec:deepcfnade}
So far we have described \cfnade with single hidden layer. As suggested
by the recent and impressive success of deep neural
networks~\citep{krizhevsky2012imagenet,szegedy2014going,he2015deep},
extending \cfnade to a deep, multiple hidden layers architecture could
allow us to have better performance. Recently, \citet{Uria2013b}
proposed an efficient deep extension to original
NADE~\citep{larochelle2011neural} for binary vector observations,
which inspires other related deep model~\citep{zheng15deep}. Following
\citep{Uria2013b}, we propose a deep variant of \cfnade.

As mentioned in Section~\ref{sec:model}, a different \cfnade model is
used for each user and the ordering $o$ in $\bf r$ is stochastically
sampled from the set of permutations of $(1,2,\ldots,D)$. Training
\cfnade on stochastically sampled orderings corresponds, in
expectation, to minimizing the cost in Equation~\ref{eqn:hybrid_cost}
over all possible orderings for each user. As noticed by
\citet{Uria2013b} and \citet{zheng15deep}, training over all possible
orderings for \cfnade implies that for any given context
${\bf r}_{{m}_{o_{<i}}}$, the model performs equally well at
predicting all the remaining items in ${\bf r}_{{m}_{o_{\ge i}}}$,
since for each item there is an ordering such that it appears at
position $i$. This is the key
observation to extend \cfnade to a deep model. Specifically, instead
of sampling a complete ordering over all the $D$ items, we instead
sample a context ${\bf r}_{{ m}_{o_{<i}}}$ and perform an update of
the conditionals using that context.

The procedure is done as follows. Given a user who has rated $D$
items, an ordering $o$ is first sampled randomly from the set of
permutations of $(1,2,\ldots,D)$ for each update and a vector
$\mathbf{r} = (r_{m_{o_1}}, r_{m_{o_2}}, \ldots, r_{m_{o_D}})$ is
generated according to the ordering $o$. Then a split point $i$ is
randomly drawn from $\{1,2,\dots,D\}$ for each update. The split point
$i$ divides ${\bf r}$ into two parts: ${\bf r}_{{m}_{o_{<i}}}$ and
${\bf r}_{{m}_{o_{\ge i}}}$. According to the analysis above, in the
new training procedure, ${\bf r}_{{m}_{o_{<i}}}$ is considered as the
input of \cfnade and the training objective is to maximize
conditionals $p(r_{m_{o_j}}|{\bf r}_{m_{o_{<i}}})$ for each element in
${\bf r}_{{m}_{o_{\ge i}}}$. The cost function with this procedure is
\begin{equation}
\cost = \frac{D}{D-i+1}\sum_{j\ge i} -\log p\left(r_{m_{o_j}}|{\bf r}_{m_{o_{<i}}}\right).
\label{eqn:deep_cost}
\end{equation}
By Equation~\ref{eqn:deep_cost}, the model predicts the ratings of
each items after the splitting position $i$ in the randomly drawn
ordering $o$ as if it were actually at position $i$. The factors in
front of the sum come from the fact that the total number of elements
in the sum is $D$ and that we are averaging over $D-i+1$ possible
choices for the item at position $i$, similar to \citep{Uria2013b} and
\citep{zheng15deep}. Derivation of Equation~\ref{eqn:deep_cost} can be
found in the supplementary materials.
%where $p(r_{m_{o_j}}|{\bf r}_{m_{o_{<i}}})$ can be computed by either Equation~\ref{eqn:cfnade_softmax} or Equation~\ref{eqn:ordinal_conditional}.

In this procedure, a training update relies only on a single hidden
representation $\mathbf{h}( {\bf r}_{m_{o_{<i}}} )$, more hidden
layers can be added into \cfnade with the
computational complexity increased moderately. Particularly, suppose
$\mathbf{h}^{(1)} ( {\bf r}_{m_{o_{<i}}} )$ is the hidden
representation computed in Equation~\ref{eqn:cfnade_hidden}. Then new hidden layers can be
added as in a regular deep feed-forward neural network:
\begin{equation}
    \mathbf{h}^{(l)}\left ( {\bf r}_{m_{o_{<i}}} \right ) = {\bf g}\left({\bf c}^{(l)} + {\bf W}^{(l)}{\bf h}^{(l-1)}\left( {\bf r}_{m_{o_{<i}}}  \right)\right)
    \label{eqn:deep_cfnade_hidden}
\end{equation}
for $l = 2,\ldots,L$, where $L$ is the total number of hidden
layers. Then the conditionals
$p( r_{m_{o_j}}=k|{\bf r}_{m_{o_{<i}}} )$, either in
Equation~\ref{eqn:cfnade_softmax} or
Equation~\ref{eqn:ordinal_conditional}, can be computed from
$\mathbf{h}^{(L)} ( {\bf r}_{m_{o_{<i}}} )$. To this end, the number of
operations a \cfnade takes for one input is $O(K\hat{D}H+H^2L)$, as
in regular multiple layers neural networks, where $\hat{D}$ is the
average number of ratings for a user and $H$ is the number of hidden
units for each layer.  We denote $\tilde{\cost}_{\textup{reg}}$ and
$\tilde{\cost}_{\textup{ord}}$ as the cost functions of
Equation~\ref{eqn:deep_cost} associated with conditionals computed by
Equation~\ref{eqn:cfnade_softmax} and
Equation~\ref{eqn:ordinal_conditional}, respectively.

Finally, similar to Equation~\ref{eqn:hybrid_cost}, we can also define a hybrid cost $\tilde{\cost}_{\textup{hybrid}}$ as 
\begin{equation}
\tilde{\cost}_{\textup{hybrid}} = (1-\lambda)\tilde{\cost}_{\textup{reg}} + \lambda\tilde{\cost}_{\textup{ord}}
\label{eqn:deep_hybrid}
\end{equation}

Note that \cfnade with a single hidden layer can also be trained by
Equation~\ref{eqn:deep_cost}. In practice,
Equation~\ref{eqn:deep_cost} can be implemented efficiently on GPUs,
hence we use it throughout our experiments for either one hidden layer
or multiple-layers architecture.

\section{Experiments}
\label{sec:experiments}

In this section, we test the performance of \cfnade on 3 real-world
benchmarks: MovieLens 1M, MovieLens 10M~\citep{harper2015movielens}
and Netflix dataset~\citep{bennett2007netflix}, which contain $10^6$,
$10^7$ and $10^8$ ratings, respectively. Following
LLORMA~\citep{lee2013local} and AutoRec~\citep{sedhain2015autorec},
$10\%$ of the ratings in each of these datasets are randomly selected
as the test set, leaving the remaining $90\%$ of the ratings as the
training set. Among the ratings in the training set, $5\%$ are used as
validation set. We use a default rating of $3$ for items without
training observations. Prediction error is measured by Root Mean
Squared Error (RMSE), %$\textit{RMSE}=\sqrt{\frac{\sum_{i=1}^N (r_i-\tilde{r}_i)^2}{N}}$
\begin{equation}
	\textit{RMSE}=\sqrt{\frac{\sum_{i=1}^S (r_i-\tilde{r}_i)^2}{S}}
	\label{eqn:rmse}
\end{equation}
where $r_i$ is the $i$\textsuperscript{th} true rating and
$\tilde{r}_i$ is the predicted rating by the model, $S$ is the total
number of ratings in the testset. We report the average RMSE on test
set over $5$ different splits and compare \cfnade with some strong
baselines including LLORMA, AutoRec, and other competitive
methods. Experimental results show that \cfnade outperforms the
state-of-the-art performance on these benchmarks.

\subsection{Datasets Description}
\label{sec:datasets}
MovieLens 1M dataset contains around $1$ million anonymous ratings of
approximately $3900$ movies by $6040$ users, where each user rated at
least $20$ items. The ratings in MovieLens 1M dataset are made on a
$5$-star scale, with $1$-star increments. MovieLen 10M dataset
contains about $10$ million ratings of $10681$ movies by $71567$
users. The users of MovieLens 10M dataset are randomly chosen and each
user rated at least $20$ movies. Unlike MovieLens 1M dataset, the
ratings in MovieLens 10M are on a $5$-star scale with
\textit{half}-star increments. Thus, the number of rating scales of
MovieLens 10M is actually $10$. In our experiments, we rescale the
ratings in MovieLens 10M to $10$-star scale with $1$-star
increments. Netflix dataset comes from the Netflix challenge
prize\footnote{The test set of Netflix prize challenge dataset is not
  available now. Following \citet{lee2013local} and
  \citet{sedhain2015autorec}, we split the available trainset of
  Netflix dataset into train, valid and test sets.}. It
is massive compared to the previous two, which contains more than
$100$ million ratings of $17770$ movies by $480189$ users. The ratings
of Netflix dataset are on $5$-star scale, with $1$-star increments.

\subsection{Experiments on MovieLen 1M Dataset}
\label{sec:movielens1m}

In this section, we test the performance of \cfnade on MovieLen 1M
dataset. We first evaluate the performance of the ordinal cost
described in Section~\ref{sec:cost} with/without sharing parameters
between different ratings as described in Section~\ref{sec:accu}. Then
we compare several variants of \cfnade with some strong baselines.

\subsubsection{The performance of the ordinal cost}
\label{sec:variants}

In this section, we evaluate the impact of the ordinal weight $\lambda$ in
Equation~\ref{eqn:deep_hybrid} on the performance of \cfnade. As
\citet{sedhain2015autorec} mentioned, item-based CF outperforms
user-based CF, therefore we use item-based \cfnade (I-\cfnade) in this
section. Distinct from user-based \cfnade (U-\cfnade), which builds a
different model for each user as we described previously, I-\cfnade
model builds a different \cfnade model for each item. In other words,
the only difference between U-\cfnade and I-\cfnade is that the roles
of users and items are switched. Comparison between U-\cfnade and
I-\cfnade can be found in Section~\ref{sec:1mbaselines}.

The configuration of the experiments is as follows. We use a single
hidden layer architecture and the number of hidden units is set to
$500$, same as AutoRec~\citep{sedhain2015autorec} and
LLORMA~\citep{lee2013local}. Adam~\citep{kingma2014adam} with default
parameters ($b_1=0.1$,$b_2=0.001$ and $\epsilon=10^{-8}$) are utilized
to optimize the cost function in Equation~\ref{eqn:deep_cost}. The
learning rate is set to $0.001$ , the weight decay is set to
$0.015$ and we use the tanh activation function. 

\begin{figure}[h]
 \begin{center}
 \centerline{\includegraphics[width=\columnwidth]{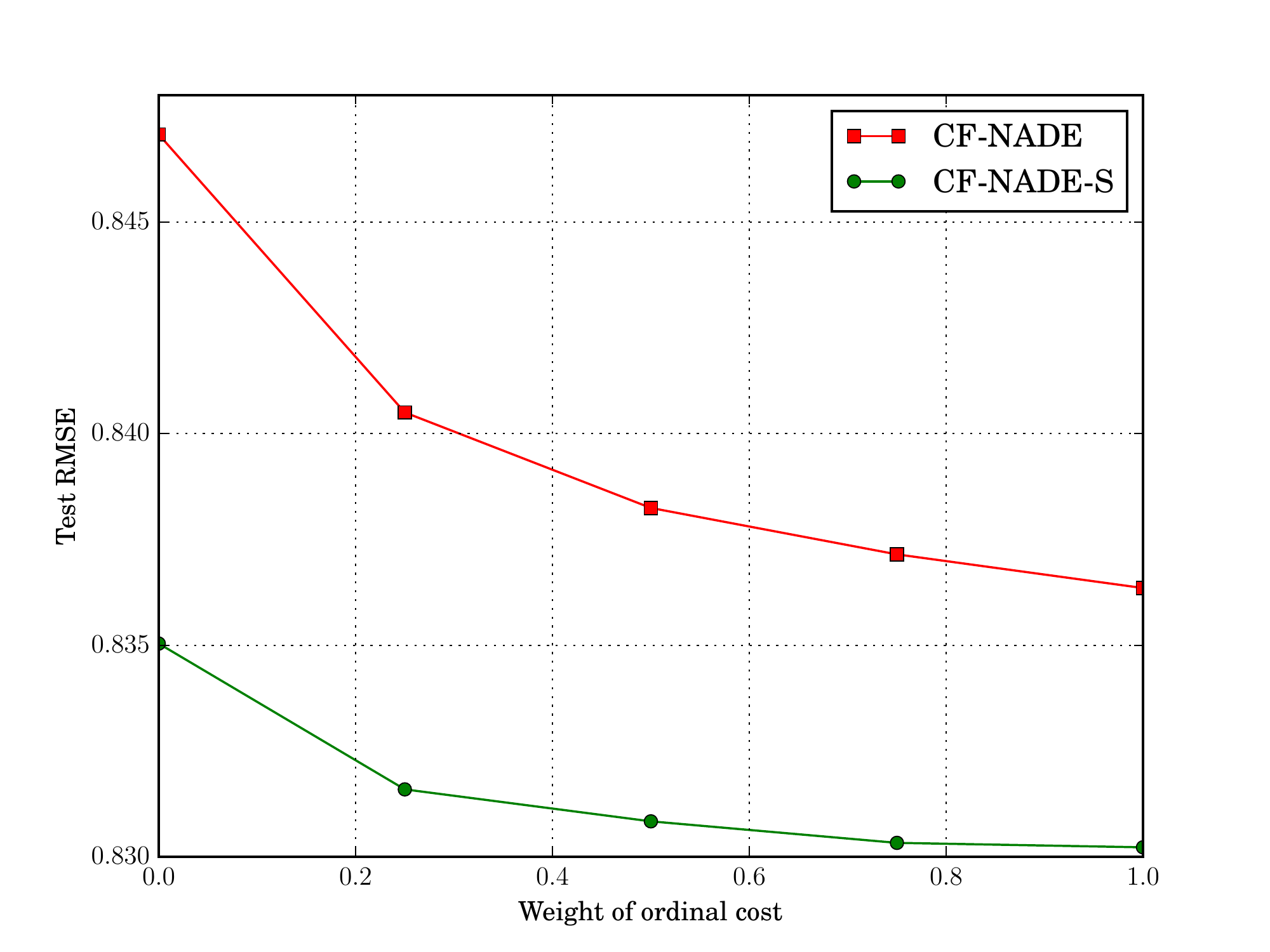}}
 \caption{The performance of \cfnade and {\cfnade}-S w.r.t ordinal weight $\lambda$ on MovieLens 1M dataset.}
 \label{fig:exp1}
 \end{center}
 \end{figure}

Let {\cfnade}-S denote the variant of \cfnade model where parameters are
shared between different ratings, as described in
Section~\ref{sec:accu}. Figure~\ref{fig:exp1} shows the superior
performance of \cfnade and {\cfnade}-S w.r.t different values of
$\lambda$. Effectiveness of parameter sharing and ordinal cost can be
justified by observing that: 1) {\cfnade}-S always outperforms regular
\cfnade; 2) as the ordinal weight $\lambda$ increases, test RMSE of
both \cfnade and {\cfnade}-S decrease monotonically. Based on these
observations, we will use {\cfnade}-S and fix $\lambda=1$ throughout
the rest of the experiments.

 \subsubsection{Comparing with strong baselines on MovieLens 1M}
 \label{sec:1mbaselines}
 
 % In this section, we will compare \cfnade with other baselines on
 % MovieLens 1M dataset.

 In this comparison, we compare \cfnade with other baselines on MoiveLens 1M dataset. During the comparison, the learning rate is chosen on the validation set by
 cross-validation among $\{0.001,0.0005,0.0002\}$, and the weight decay is
 chosen among $\{0.015,0.02\}$.  According to
 Section~\ref{sec:variants}, the weight $\lambda$ of ordinal cost is
 fixed to $1$ and {\cfnade}-S is adopted. The model is trained with
 Adam optimizer and tanh as activation function.
 
 \begin{figure}[h]
 \begin{center}
 \centerline{\includegraphics[width=\columnwidth]{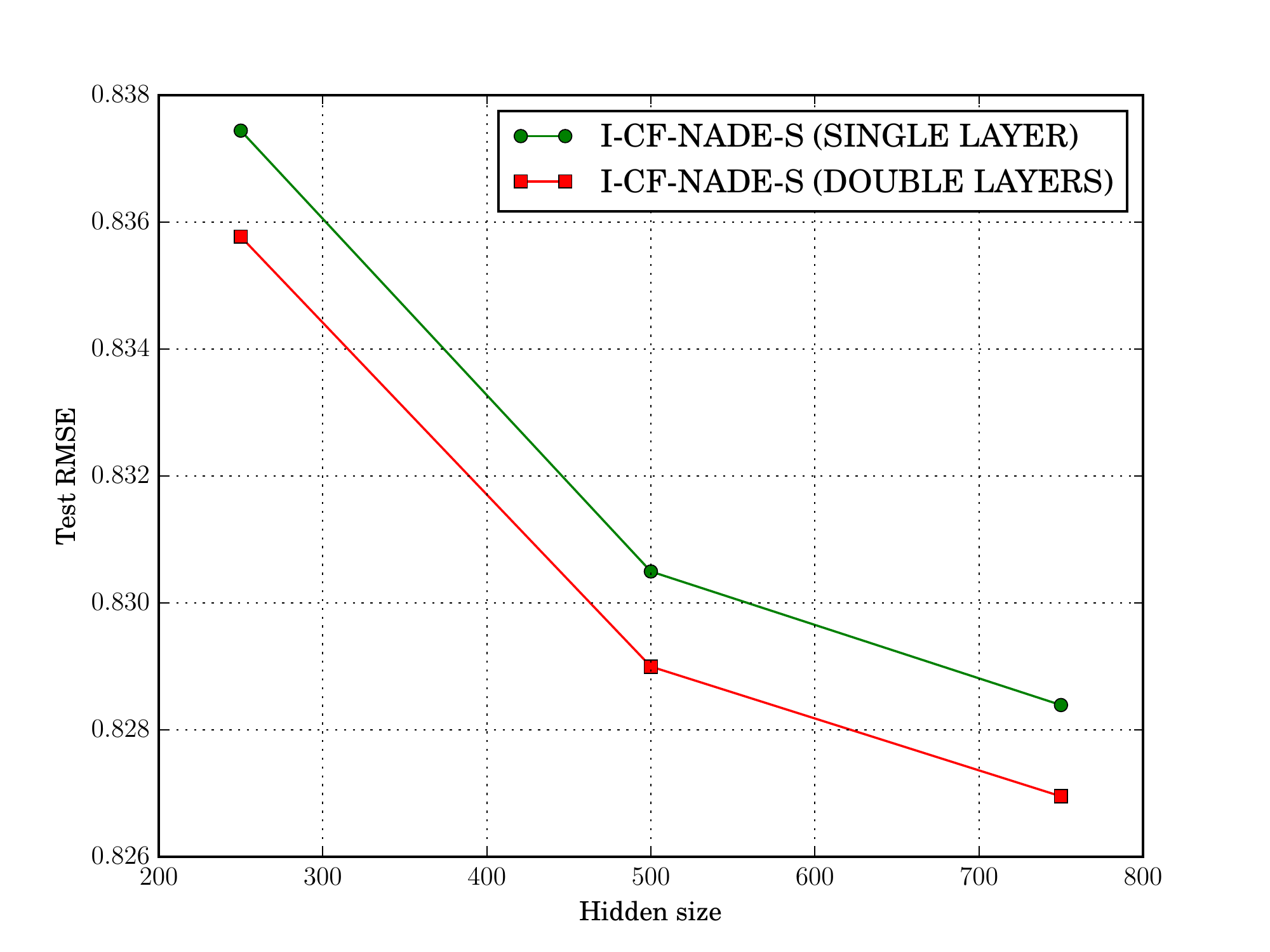}}
 \caption{The performance of I-\cfnade w.r.t the number of hidden units on MovieLens 1M dataset.}
 \label{fig:exp2_a}
 \end{center}
 \end{figure}
 
 Table~\ref{tab:movielens1m} shows the performance of {\cfnade}-S and
 baselines. The number of hidden units of \cfnade is $500$, same as
 AutoRec~\citep{sedhain2015autorec} for a fair comparison. One can
 observe that I-{\cfnade}-S outperforms U-{\cfnade}-S by a
 large margin. I-{\cfnade}-S with a single hidden
 layer achieves RMSE of $0.830$, which is comparable with any strong
 baseline. Moreover, I-{\cfnade}-S with 2 hidden layers achieves RMSE of $0.829$.

 Figure~\ref{fig:exp2_a} illustrates the performance of I-{\cfnade}-S
 w.r.t the number of hidden units. Increasing the number of hidden
 units is beneficial, but the return is
 diminishing. It can also be observed from Figure~\ref{fig:exp2_a} that deep \cfnade models
 achieve better performance than the shallow ones, as expected.

\begin{table}[h]
 \caption{Test RMSE of different models on MovieLens 1M.}
 \label{tab:movielens1m}
 \begin{center}
 \begin{small}
 \begin{sc}
 \begin{tabular}{lc}
 \hline
 \abovespace\belowspace
 Method & Test RMSE  \\
 \hline
 \abovespace
 PMF$\dagger$    & 0.883 \\
 U-RBM$\ast$ & 0.881\\
 U-AutoRec~\citep{sedhain2015autorec} & 0.874\\
 LLORMA-Global~\citep{lee2013local} & 0.865\\
 I-RBM$\ast$ & 0.854\\
 BiasMF$\ast$  & 0.845\\
 NNMF~\citep{dziugaite2015neural} &0.843\\
 LLORMA-Local~\citep{lee2013local} & 0.833\\
 \belowspace
 I-AutoRec~\citep{sedhain2015autorec} & 0.831\\
 U-{\cfnade}-S (single layer)  & 0.850 \\
 U-{\cfnade}-S (2 layers )  & 0.845 \\
 I-{\cfnade}-S (single layer) & 0.830\\
 I-{\cfnade}-S (2 layers) & {\bf 0.829}\\
 \hline
 \end{tabular}
 \end{sc}
 \end{small}
 \begin{minipage}{0.48\textwidth}
{\small
$\dagger$: Taken from~\citep{dziugaite2015neural}. \\
$\ast$: Taken from~\citep{sedhain2015autorec}.
}
\end{minipage}
 \end{center}
 \end{table}

\subsection{Experiments on MovieLens 10M Dataset}
\label{sec:movielens10m}
% We then conduct experiments of \cfnade on the MovieLens 10M dataset.
As mentioned in Section~\ref{sec:datasets}, the MovieLens 10M 
% contains $10681$ movies and $71567$ users, which
is much bigger than the MovieLens 1M, so we opt to use the factored
version of \cfnade described in Section~\ref{sec:factored} and set
$J=50$. Even in this setting, I-\cfnade with $71567$ users and $10$
rating scales will still bring about as many as $70$ million free
parameters,
% even the factored \cfnade is used and with $J=50$, which might be
% problematic.
Hence, we only report the performance of
U-\cfnade in this experiment. Same as in Section~\ref{sec:variants},
we train the model with Adam optimizer and using $\mathrm{tanh}$ as activation
function.  Other configurations are as follows: The number of hidden
units to $500$, the weight decay is $0.015$ and $\lambda$ is set to
$1$ and parameters are shared between ratings following
Section~\ref{sec:variants}. The base learning rate is $0.0005$, and we
double it for the parameters of the first layer.

Table~\ref{tab:movielens10m} shows the comparison between \cfnade and
other baselines on MovieLens 10M dataset. U-{\cfnade}-S with a single
hidden layer has already outperformed the baselines, which achieves
RMSE of $0.772$. The performance of U-{\cfnade}-S can be slightly
improved by adding another hidden layer. Noticeably, the test RMSE of
U-AutoRec is much worse than I-AutoRec, whereas U-{\cfnade}-S
outperforms I-AutoRec.

\begin{table}[h]
 \caption{Test RMSE of different models on MovieLens 10M.}
 \label{tab:movielens10m}
 \begin{center}
 \begin{small}
 \begin{sc}
 \begin{tabular}{lc}
 \hline
 \abovespace\belowspace
 Method & Test RMSE  \\
 \hline
 \abovespace
 U-AutoRec~\citep{sedhain2015autorec} & 0.867\\
 I-RBM$\dagger$ & 0.825\\
 U-RBM$\dagger$ & 0.823\\
 LLORMA-Global~\citep{lee2013local} & 0.822\\
 BiasMF$\dagger$ & 0.803 \\
 LLORMA-Local~\citep{lee2013local} & 0.782\\
 \belowspace
 I-AutoRec~\citep{sedhain2015autorec} & 0.782\\
 U-{\cfnade}-S (single layer) & 0.772\\
 U-{\cfnade}-S (2 layers) & {\bf 0.771}\\
 \hline
 \end{tabular}
 \end{sc}
 \end{small}
 \begin{minipage}{0.48\textwidth}
{\small
$\dagger$: Taken from~\citep{sedhain2015autorec}.
}
\end{minipage}
 \end{center}
 \end{table}
 
 \subsection{Experiments on Netflix Dataset}
 \label{sec:netflix}
 Our final set of experiments are on the massive Netflix dataset,
 which contains $10^8$ ratings. Similar to
 Section~\ref{sec:movielens10m}, we use the factored version of
 U-\cfnade with $J=50$. The Netflix dataset is so big that we need not
 add a strong regularization to avoid overfitting and therefore set
 the weight decay to $0.001$. Other configurations are the same as in
 Section~\ref{sec:movielens10m}.
 
 Table~\ref{tab:netflix} compares the performance of U-\cfnade with
 other baselines. We can see that U-{\cfnade}-S with a single
 hidden layer achieves RMSE of $0.804$, outperforming all
 baselines. Another observation from Table~\ref{tab:netflix} is that
 using a deep \cfnade architecture achieves a slight improvement over
 the shallow one, with a test RMSE of $0.803$.
 
 \begin{table}[h]
 \caption{Test RMSE of different models on Netflix dataset.}
 \label{tab:netflix}
 \begin{center}
 \begin{small}
 \begin{sc}
 \begin{tabular}{lc}
 \hline
 \abovespace\belowspace
 Methods & Test RMSE  \\
 \hline
 \abovespace
 LLORMA-Global~\citep{lee2013local} & 0.874\\
 U-RBM$\dagger$ & 0.845\\
 BiasMF$\dagger$ & 0.844 \\
 LLORMA-Local~\citep{lee2013local} & 0.834\\
 \belowspace
 I-AutoRec~\citep{sedhain2015autorec} & 0.823\\
 U-{\cfnade}-S (single layer) & 0.804\\
 U-{\cfnade}-S (2 layers) & {\bf 0.803}\\
 \hline
 \end{tabular}
 \end{sc}
 \end{small}
 \begin{minipage}{0.48\textwidth}
{\small
$\dagger$: Taken from~\citep{sedhain2015autorec}.
}
\end{minipage}
 \end{center}
 \end{table}
 
 \begin{table}[h]
\begin{small}
\centering
\caption{Complexity of \cfnade on different benchmarks}
\label{tab:complexity}
\begin{tabular}{@{}lccccc@{}}
\toprule
                Dataset & $\#$Layers & $\#$Params & Train Time & Test Time\\ 
                &&(million)&(second)&(second)\\ \midrule
\multirow{2}{*}{ML 1M} &  $1$   & $30.2$ &$3.09$   & $0.65$  \\
                  & $2$ & $30.48$ & $3.11$ & $0.68$ \\
\multirow{2}{*}{ML 10M} & $1$ & $10.78$ & $134.76$ & $31.72$ \\
                  & $2$ & $10.98$ & $135.62$ & $32.73$ \\
\multirow{2}{*}{Netflix} & $1$ & $9.02$  & $1057.81$ & $239.78$ \\
                  & $2$ & $9.19$ &  $1064.33$& $243.79$ \\ \bottomrule
\end{tabular}
\end{small}
\end{table}

%3.09	0.65	134.76	31.72	1057.81	239.78
%3.11	0.68	135.62	32.73	1064.33	243.79
 
\subsection{The Complexity and Running Time of CF-NADE}
\label{exp:complexity}

%As mentioned in Sec.~\ref{sec:deepcfnade}, the complexity of \cfnade is $O(\hat{D}H+H^2N)$. 

We implement \cfnade using Theano~\cite{Bastien-Theano-2012} and
Blocks~\citep{van2015blocks}, and the code is available at
\url{https://github.com/Ian09/CF-NADE}. Table~\ref{tab:complexity} shows
the running time of one epoch\footnote{Experiments are conducted on a single NVIDIA Titan X Card.}  as
well as the number of parameters used by \cfnade. For MovieLens 1M dataset, we used the
item-based \cfnade and did not use the factorization method introduced
by Sec~\ref{sec:factored}, hence the number of parameters for
MovieLens 1M is bigger than the other two. Running times in
Table~\ref{tab:complexity} include overheads such as transferring data
from and to GPU memory for each update. Note that
there is still room for faster implementations\footnote{In our implementation, samples are
  represented as $M\times K$ binary matrices, where $M$ is the number
  of items and $K$ is the number of rating scales. An entry $(m,k)$ is
  assigned $1$ only if the user gave a $k$-star to item $m$. Thus, we
  could use the {\it tensordot} operator in Theano and feed \cfnade
  with a batch of samples. In the experiments, mini-batch size is set
  to $512$. One disadvantage of this implementation is that some
  amount of computational time is spent on unrated items, which can be
  enormous especially when the data is sparse.
%Hence, running time in
%  Table~\ref{tab:complexity} could be improved with more efficient
%  implementation. 
}.
% given that the complexity of the model is
%$O(\hat{D}H+H^2L)$ (c.f. Sec~\ref{sec:deepcfnade}).

%We list the running time and the number of parameters in TableXX. 

\section{Conclusions}
\label{sec:conclusion}

In this paper, we propose \cfnade, a feed-forward, autoregressive
architecture for collaborative filtering tasks. \cfnade is inspired by
the seminal work of RBM-CF and the
recent advancements of NADE. We 
propose to share parameters between different ratings to improve the
performance. We also describe a factored version of
\cfnade, which reduces the number of parameters by factorizing a large matrix by a product of two lower-rank matrices,
for better scalability. Moreover, we take the ordinal nature of
preference into consideration and propose an ordinal cost to optimize \cfnade. Finally,
following recent advancements of deep learning, we extend \cfnade to a
deep model with moderate increase of computational
complexity. Experimental results on three real-world benchmark datasets show
that \cfnade outperforms the state-of-the-art methods on
collaborative filtering tasks. all results of this work rely on explicit feedback, namely, ratings explicitly given by users. however, explicit feedback is not always available or as common as implicit feedback (watch, search, browse behaviors) in real-world recommender systems~\cite{hu2008collaborative}. Developing a version of \cfnade tailored for implicit feedback is left for future work.

\section*{Acknowledgements} 
We thank Hugo Larochelle and the reviewers for many helpful discussions.

\bibliography{example_paper}
\bibliographystyle{icml2016}

\end{document}